\documentclass[aps,prl,reprint,showpacs,superscriptaddress]{revtex4-1}
\usepackage{verbatim}

\usepackage{graphicx}
\usepackage{dcolumn}
\usepackage{bm}
\usepackage{color}
\usepackage{lipsum}
\usepackage{amsmath}
\usepackage{mathrsfs}
\usepackage{lineno}
\usepackage[colorlinks,urlcolor=blue,linkcolor=blue,citecolor=blue]{hyperref}
\usepackage{siunitx}
\usepackage{multirow}
\begin{document}


\title{Colliding of two high Mach-number quantum degenerate plasma jets}
\author{W.-B. Zhang}
\affiliation{Key Laboratory for Laser Plasmas and Department of Physics and Astronomy, Shanghai Jiao Tong University, Shanghai 200240, People’s Republic of China}
\affiliation{Collaborative Innovation Center of IFSA (CICIFSA), Shanghai Jiao Tong University, Shanghai 200240, People’s Republic of China}
\author{Y.-H. Li}
\affiliation{Institute of Physics, Chinese Academy of Sciences, Beijing 100190, People’s Republic of China}
\affiliation{University of Chinese Academy of Sciences, Beijing 100049, People’s Republic of China}
\affiliation{Collaborative Innovation Center of IFSA (CICIFSA), Shanghai Jiao Tong University, Shanghai 200240, People’s Republic of China}
\author{D. Wu}
\email{dwu.phys@sjtu.edu.cn}
\affiliation{Key Laboratory for Laser Plasmas and Department of Physics and Astronomy, Shanghai Jiao Tong University, Shanghai 200240, People’s Republic of China}
\affiliation{Collaborative Innovation Center of IFSA (CICIFSA), Shanghai Jiao Tong University, Shanghai 200240, People’s Republic of China}
\author{J. Zhang}
\email{jzhang@iphy.ac.cn}
\affiliation{Institute of Physics, Chinese Academy of Sciences, Beijing 100190, People’s Republic of China}
\affiliation{University of Chinese Academy of Sciences, Beijing 100049, People’s Republic of China}
\affiliation{Key Laboratory for Laser Plasmas and Department of Physics and Astronomy, Shanghai Jiao Tong University, Shanghai 200240, People’s Republic of China}
\affiliation{Collaborative Innovation Center of IFSA (CICIFSA), Shanghai Jiao Tong University, Shanghai 200240, People’s Republic of China}

\date{\today}
\begin{abstract}
Colliding of two high Mach-number quantum degenerate plasmas is one of the most essential components in the double-cone ignition (DCI) inertial confinement fusion scheme, in which two highly compressed plasma jets from the cone-tips collide along with rapid conversion from the colliding kinetic energies to the internal energy of a stagnated isochoric plasma. Due to the effects of high densities and high Mach-numbers of the colliding plasma jets, quantum degeneracy and kinetic physics might play important roles and challenge the predictions of traditional hydrodynamic models. In this work, the colliding process of two high Mach number quantum degenerate Deuterium-plasma jets with sizable scale ($\sim 1000\ \si{\mu m}$, $\sim 300\ \si{ps}$, $\sim 100\ \si{g/cc}$, $\sim 300\ \si{km/s}$) were investigated with first-principle kinetic simulations and theoretical analyses. In order to achieve high-density compression, the colliding kinetic pressure should be significantly higher than the pressure raised by the quantum degeneracy. This means high colliding Mach numbers are required. However, when the Mach number is further increased, we surprisingly found a decreasing trend of density compression, due to kinetic effects. It is therefore suggested that there is theoretically optimal colliding velocity to achieve the highest density compression. Our results would provide valuable suggestions for the base-line design of the DCI experiments and also might be of relevance in some violent astrophysical processes, such as the merger of two white dwarfs.
\end{abstract}


\maketitle

\begin{figure}
\includegraphics[width=8.5cm]{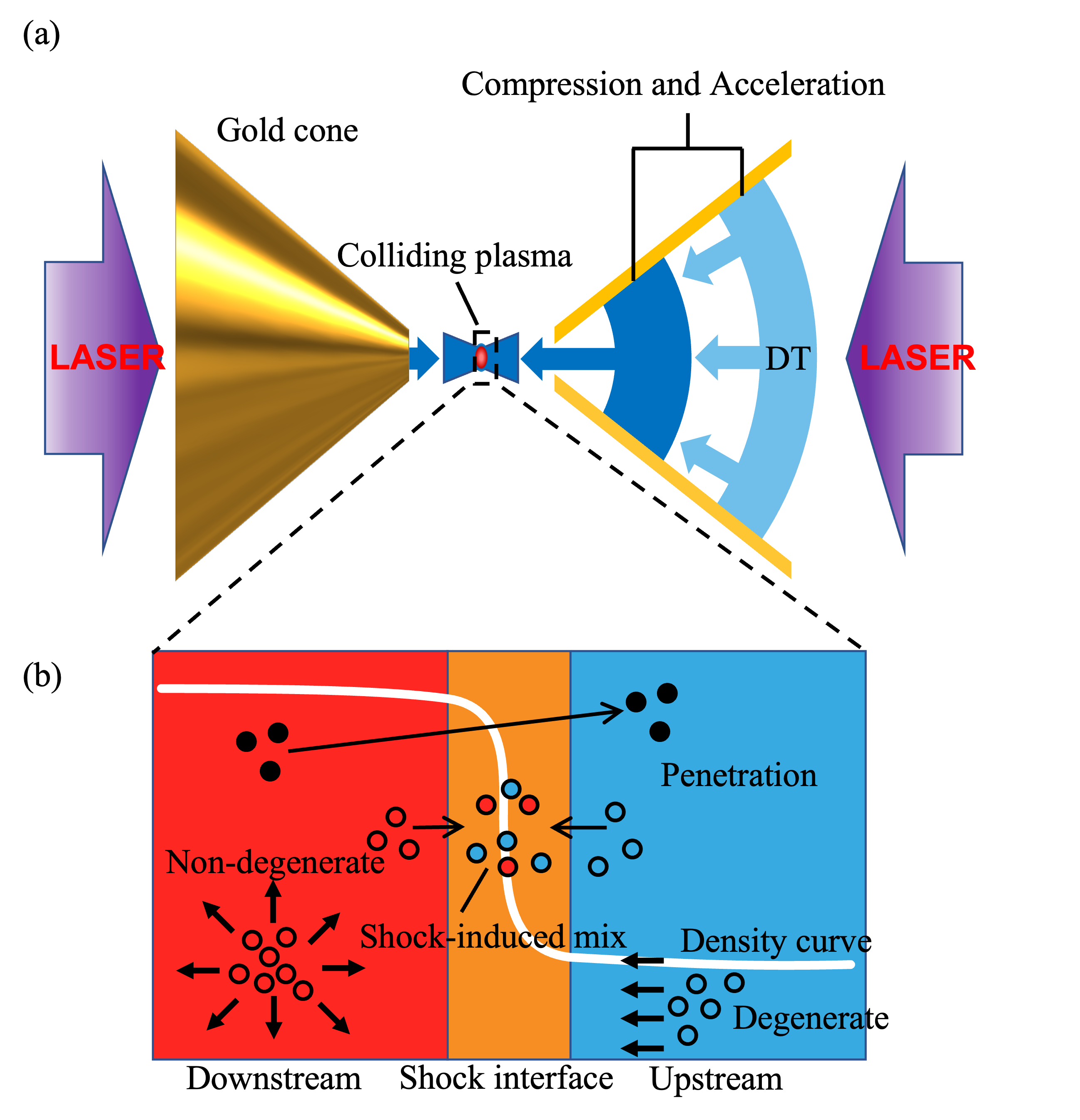} 
\caption{\label{fig1}(color online). (a) Schematic of head-on collision in the DCI scheme. The fuels are initially compressed and accelerated by lasers. Then two plasma jets eject out from gold cone-tips and collide with each other with high Mach number, forming strong shocks in the colliding region. (b) Key features near the shock front: the blue, orange and red regions represent the upstream, shock interface and downstream respectively. Shock is maintained through the equilibrium between kinetic and thermal pressure. Degenerate and non-degenerate plasmas mix in the shock interface, resulting in non-equilibrium states at the shock front. The high energy ions in the downstream region penetrate upstream beyond the shock front, leading to enhancement of the shock width \cite{Keenan2017pre}.} 
\end{figure}

Double-cone ignition (DCI) \cite{Zhang2020DoubleconeIS} is a new type of laser inertial confinement fusion (ICF) scheme proposed recently. In this scheme, the deuterium-tritium (DT) fuel shells assembled in two head-on gold cones are compressed and accelerated along the cone axis by carefully tailored nanosecond laser pulses, forming high-speed DT plasma jets ($\sim 100\ \si{g/cc}$, $\sim 300\ \si{km/s}$) from the cone-tips, and then collide with each other in the central open space (shown in Fig.\ \ref{fig1}(a)). Due to the momentum filtering and transverse confinement of the wall \cite{Zhang2023Nature}, the colliding DT fuels remain cryogenic ($\sim 50\ \si{eV}$) and fall in highly degenerated states. During the colliding process, densities of the fuels increase rapidly and reach the required $\sim 300\ \si{g/cc}$ \cite{Atzeni1999InertialFF}. Finally, fast electrons generated by picosecond petawatt laser pulses are injected into the stagnated isochoric plasma perpendicular to the colliding direction, locally heating plasma to $\si{keV}$s.

The colliding of two highly compressed DT plasma jets from the cone-tips is one of the key components in the DCI scheme. Distinguished from the conventional ICF schemes where the DT pellet undergoes a spherical stagnation \cite{Nuckolls1972LaserCO}, a strong colliding shock, as depicted in Fig.\ \ref{fig1}(a), is generated to convert the kinetic energies of the DT plasma jets to their internal energies, forming an isochoric preheated plasma in the colliding center for the following fast heatings.

The colliding of two plasma jets is also an active research area in ICF and laboratory astrophysics communities. Previous studies mainly focused on either collisionless kinetic shocks \cite{Takabe2008HighMachNC,Sorasio2005VeryHM,Kuramitsu2011TimeEO,Boella2017InteractionBE,Kato2010ElectrostaticAE} or collisional hydrodynamic shocks \cite{Hu1972CollisionalTO,Ghosh2002IonAS,Adak2016MagnetosonicSW,Passalidis2020HydrodynamicCM}. In indirect-drive ICF schemes, collisionless shock appears when the high-Z plasma expands from the hohlraum wall and collides with filling gases or blow-off from the fuel capsule \cite{Cai2020StudyOT,Shan2018ExperimentalEO,Zhang2017AnomalousNY}. As for the collisional cases, a representative research topic is the shock ignition scheme \cite{betti2007shockIO,lafon2010gainCA,sauppe2016simulationsSO}, in which the shocks are produced and maintained in high density DT fuels. In recent laboratory astrophysics studies, the oblique merge of supersonic plasma jets \cite{Merritt2013ExperimentalCO}, interpenetration and stagnation of colliding plasma \cite{Chenaispopovics1997KineticTT,AlShboul2014InterpenetrationAS} and shock-generated electromagnetic fields \cite{Hua2019PhysRevLett} were also investigated experimentally.

With in-depth research, it is found that conventional hydrodynamic theory is inadequate to describe strong shocks propagating in plasmas, especially for the case of high Mach numbers \cite{Rambo1994InterpenetrationAI}. As a result, kinetic approaches \cite{MottSmith1951TheSO,Rosenbluth1957,Tidman1958,Abe1975StrongPS} of shock front and series of new PIC and hydro-PIC hybrid simulation methods \cite{sentoku2008numericalMF,sagert2014hydrodynamicSW,lembege2001hybridAP,cohen2010simulationOL,Cai2021HybridFM} have been successively developed. Researcher's attention was shifted to kinetic effects in the ablation shock wave during the implosion and compression process in ICF, including the electron thermal conduction \cite{zhang2022influenceOT}, species separation of deuterium and tritium \cite{PhysRevE.91.023103,PhysRevE.90.013101,Bellei2014SpeciesSA}, and non-local transport in the shock front \cite{Rosenberg2014ExplorationOT,Albright2013RevisedKR}. Recently, the quantum hydrodynamic method has also been introduced \cite{Doyon2017LargeScaleDO,Simmons2020WhatIA,Graziani2021ShockPI} to describe the quantum effects in colliding shocks.

Although a great deal of work had been done to study colliding shock waves, it is still a challenge to study the colliding DT fuels in the DCI scheme. For the colliding process of the DCI scheme, significant non-equilibrium phenomena exist near the shock interface beyond hydrodynamics, including shock-induced mix and penetration of high energy ions (shown in Fig.\ \ref{fig1}(b)). On the other hand, to simulate large-scale dynamics of high-density plasmas, the numerical noise and cost of computational resources are unaffordable for most PIC codes. In the meanwhile, due to the momentum filtering and transverse confinement of the wall \cite{Zhang2023Nature}, the colliding of the two highly compressed DT plasma jets fall in highly degenerated states, which typically can not be treated as classical plasmas.

In order to tackle the above challenges, we have developed a new simulation method \cite{Wu2020PhysRevE,Wu2023} with an ingenious kinetic-ion and kinetic/hydrodynamic-electron treatment. This method takes advantage of modern particle simulation techniques and binary Monte Carlo collisions, including both long-range collective electromagnetic fields and short-range particle-particle interactions, thereby collisional coupling and state-dependent coeﬀicients, that are usually approximately used with different forms in fluid descriptions, are removed. Especially, in this method \cite{Wu2023}, the restrictions of simulation grid size and time step on electron scales, which usually appear in a fully kinetic description, are eliminated. In order to take quantum degeneracy into account, the Boltzmann-Uehling-Uhlenbeck equation is adopted for the transport of electrons, and the Fermi-Dirac distributions and the Pauli-exclusion principle among electrons are naturally fulfilled in the above first principle kinetic method.

For simplicity, we conducted large-scale one-dimensional simulations for the collidings of two pure Deuterium-plasmas (D-plasmas) to avoid the effects of ion species separation. In order to achieve high-density compression, colliding Mach number should be high enough to ensure that the colliding kinetic pressure is significantly higher than the pressure raised by the quantum degeneracy. Moreover, we surprisingly found a decreasing trend of density compression in the colliding center as the Mach number is larger than a particular value. This is rarely discussed in previous works. This work may be not only of significance to the base-line design of DCI scheme, but also instructive to astronomical studies in respect of the merger of super-dense objects such as white dwarfs \cite{Gvaramadze2019AMW}.

The configuration of our simulations is listed as follows. In the $1000\ \si{\mu m}$ simulation region, two plasma jets with sizes of $250\ \si{\mu m}$ are symmetrically set and cling to each other. The plasmas have initially uniform temperature of $50\ \si{eV}$ and central density of $100\ \si{g/cc}$. The colliding velocity is assigned from $100\ \si{km/s}$ to $900\ \si{km/s}$ at intervals of $100\ \si{km/s}$, corresponding to the Mach number $M$ ranging from $M=1.6$ to $M=14.2$. The details of the simulation parameters (time step, grid size and particles per cell) are presented in Sec.\ \uppercase\expandafter{\romannumeral1} of the Supplemental Material \cite{SM}. To ensure robustness and correctness, the convergence benchmarks with varying simulation parameters are also performed, in Sec.\ \uppercase\expandafter{\romannumeral1} of the Supplemental Material \cite{SM}. 

\begin{figure*}
\includegraphics[width=17cm]{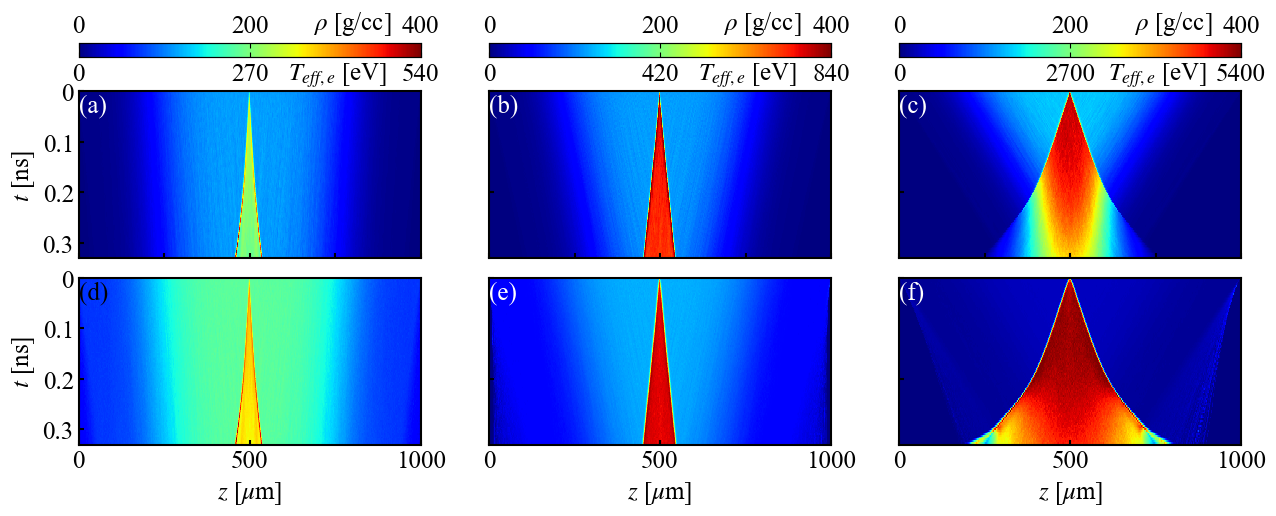} 
\caption{\label{fig2}(color online). The spatial-temporal evolutions of plasma density (in first row), effective electron temperature (in second row): (a) and (d) show the case when the colliding velocity is $v_0=100\ \si{km/s}$. (b) and (e) show the case when $v_0=300\ \si{km/s}$. (c) and (f) show the case when $v_0=900\ \si{km/s}$. Values in the color-bar represent plasma density $\rho$ in unit of $\si{g/cc}$ and effective electron temperature $T_{eff,e}$ in unit of $\si{eV}$.} 
\end{figure*}

\begin{figure}
\includegraphics[width=8.5cm]{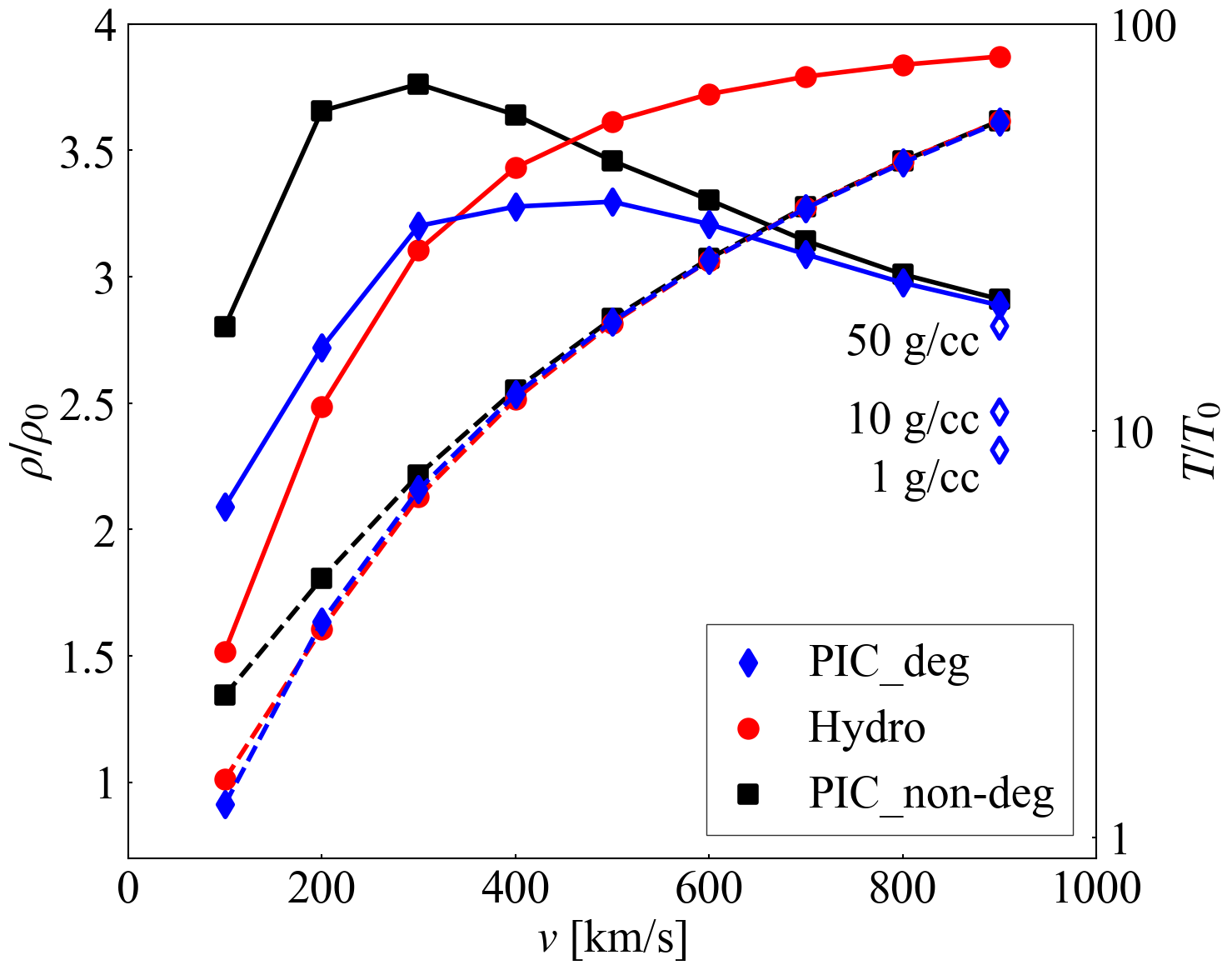} 
\caption{\label{fig3}(color online). Density (in solid lines) and temperature (in dotted lines) of the post-shock plasma, in unit of their initial values, where $\rho_0=100$ g/cc and $T_{i,e}=50$ eV. The blue diamonds represent data obtained by PIC simulations with the effects of quantum degeneracy; the red dots represent data obtained by the hydrodynamic model, and the black squares represent data obtained by classical PIC simulations. The hollow diamonds show the cases where the initial density are $\rho_0=50\ \si{g/cc}$, $\rho_0=10\ \si{g/cc}$ and $\rho_0=1\ \si{g/cc}$ respectively.} 
\end{figure}

Fig.\ \ref{fig2} shows typical simulation results of density and ``effective electron temperature'' for the colliding plasmas, where the ``effective electron temperature'' is equivalently represented by the average electron kinetic energy (the drift kinetic energy of electrons is ignorable compared with its internal energy) for convenience of counting. In the simulations, it is found that there are two shock waves propagating outward from the colliding center, characterized by sharp density and temperature discontinuities. The density and temperature of the central plasmas behind the colliding fronts are several times of that in unperturbed cold plasmas. Meanwhile, rarefaction waves enter the colliding plasmas from outside, and rapidly decrease the density and temperature of plasmas after the collision. After intersecting with the rarefaction, the shock declines and finally vanishes, as displayed in Fig.\ \ref{fig2}(c) and (f) after $t=0.2\ \si{ns}$. Nevertheless, in the early stage of the colliding, the density and electron temperature of plasmas behind shock fronts almost remain spatially and temporally uniform. This ensures the collections of average density and effective temperature of post-shock plasmas during the colliding as shown in Fig.\ \ref{fig3}.

We have made a hydrodynamic model for the colliding. According to the Rankine–Hugoniot relation, the density compression ratio $\rho_2/\rho_1$ depends on the pressure $p$ in the pre-shock (index 1) and post-shock (index 2) as follows

\begin{equation}
  \frac{\rho_2}{\rho_1}=\frac{(\gamma+1)p_2+(\gamma-1)p_1}{(\gamma+1)p_1+(\gamma-1)p_2}.
  \label{eq1}
\end{equation}

\noindent Taking the adiabatic coefficient $\gamma$ as $5/3$ for monatomic plasmas, the theoretical supremum of the density compression ratio for a single shock is 4, on condition that $p_2\ll p_1$ and $\rho_2/\rho_1=(\gamma+1)/(\gamma-1)$. 

For one-dimensional problems, we assume that the kinetic energies of ions are completely converted to their internal energies, and that thermal equilibrium is reached between electrons and ions (i.e. $T_i=T_e=T$). These assumptions are confirmed by our simulations. Consider a certain fluid element at a fixed position, the conservation of energy before and after the shock front is written as

For one-dimensional problems, we assume that the kinetic energies of ions are completely converted to their internal energies, and that thermal equilibrium is reached between electrons and ions (i.e. $T_i=T_e=T$). These assumptions are confirmed by our simulations (see Sec.\ \uppercase\expandafter{\romannumeral2} of \cite{SM}). Consider a certain fluid element at a fixed position, the conservation of energy before and after the shock front is written as

\begin{equation}
  \frac{1}{2}m_\text{D}v^2=\frac{3}{2} k_B (T_2-T_1)+[\varepsilon_e (T_2,n_2)-\varepsilon_e (T_1,n_1)],
  \label{eq2}
\end{equation}

\noindent where $m_\text{D}$ is the mass of a deuterium ion, $\varepsilon_e$ is the average energy of electrons and $n_i=\rho_i/m_D$ is the number density of D-plasmas. In Eq.\ (\ref{eq2}), D-ions lie in states that can be well described by ideal gas models, with constant heat capacity $c_v=(3/2)k_B$, and $\varepsilon_i=(3/2)k_BT$. Electrons, especially in the pre-shock regions, lie in quantum degenerate states, and follow Fermi-Dirac distributions

\begin{equation}
  f_e(E;T_e,n_e)=\frac{(2m_e)^{3/2}}{n_e \hbar^3\pi^2}\frac{\sqrt{E}}{\exp[(E/T_e)-\eta]+1},
  \label{eq3}
\end{equation}

\noindent which is normalized by $\int{f_e (E)dE}=1$ to determine the coefficient $\eta$. The effective electron temperature $\varepsilon_e(T_e,n_e)$ in Eq.\ (\ref{eq2}), is $\varepsilon_e(T_e,n_e)=\int{Ef_e(E;T_e,n_e)dE}$, which is determined by both thermal temperature $T_e$ and number density $n_e$.

It is noticed that as $T_e \to 0$, the normalizing coefficient $\eta$ in Eq.\ (\ref{eq3}) approaches $\infty$ and $\varepsilon_e$ becomes almost independent of $T_e$, being merely proportional to $n_e^{2/3}$. The energy related to the density of fermions at zero temperature is called Fermi energy. Relatively, we also performed simulations to analyze the scenario treating electrons classically, and for this scenario, $\varepsilon_e'$ also equals to $(3/2)k_BT$. Therefore $\varepsilon_e$ is always higher than $\varepsilon_e'$ for fixed electron densities and temperatures.

It can also be proved that for Fermi-Dirac distributions, the equation of state (EOS) $p_e=(2/3) n_e \varepsilon_e$ always holds, identical to the EOS of classical ideal monatomic gas. Therefore, we have

\begin{equation}
  p=\frac{2}{3}n(\varepsilon_i+\varepsilon_e).
  \label{eq4}
\end{equation}

\noindent By combining Eq.\ (\ref{eq1})-(\ref{eq4}), the post-shock density and temperature of plasmas in the colliding center can be obtained.

Fig.\ \ref{fig3} shows the densities and temperatures of the central plasmas. For simulations, the density and temperature values are picked at the early stage of the colliding, where both values remain of spatially and temporally uniform. When comparing the results between simulations with the effects of quantum degeneracy and classical models, it is found that the density of post-shock plasmas in the former is less than that in the latter, especially for low colliding velocities.

Additionally, a noteworthy aspect of Fig.\ \ref{fig3} is the comparison between the PIC simulation results and the hydrodynamic calculations for high velocities. It is observed that the post-shock density in the simulations shows an opposite trend to the theoretical calculations when $v$ is greater than $500\ \si{km/s}$, that is, $M>8$. The red lines representing the hydrodynamics predictions of densities keeps rising approaching the 4 times limit of shock compression; in contrast, the blue line of simulation results reaches a maximum ratio of $3.3$ and then decreases. The results of both degenerate and classical simulations converge when $M$ is high, indicating that the effects of quantum degeneracy are no longer significant since the colliding has heated electrons up to classical states. 

The divergence of density trends in Fig.\ \ref{fig3} has indicated that hydrodynamics is not applicable to strong shocks in the supreme high Mach number collidings. Extra simulations have been conducted with initial density of $1\ \si{g/cc}$, $10 \ \si{g/cc}$ and $50\ \si{g/cc}$ under the same colliding velocity of $v=900\ \si{km/s}$, and the results are marked on Fig.\ \ref{fig3} with open diamonds. It is noted that in the post-shock region, the compression ratio $\rho/\rho_0$ is merely $2.3$ at $\rho_0=1\ \si{g/cc}$, while rising to $2.9$ as $\rho_0$ increases to $100\ \si{g/cc}$. This result conflicts with Eq.\ (\ref{eq1}), where the compression ratio of shock is independent of the density of unperturbed pre-shock plasmas.

\begin{figure*}
\includegraphics[width=17cm]{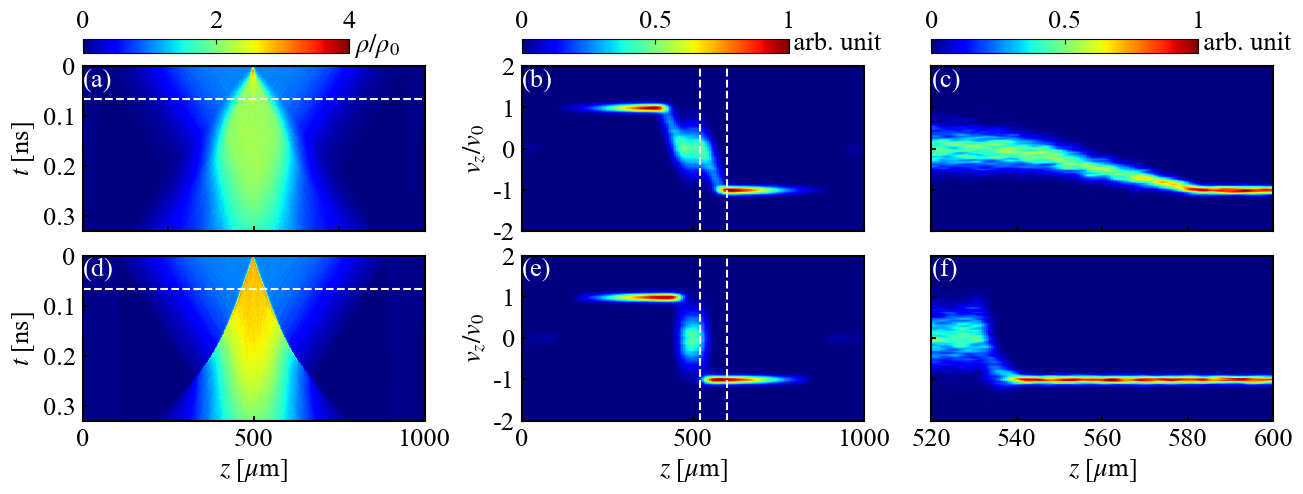} 
\caption{\label{fig4}(color online). The first row represents simulation results with initial density of $1\ \si{g/cc}$ and the second row represents simulation results with initial density of $50\ \si{g/cc}$. (a) and (d) show the spatial-temporal evolutions of plasma density with color-bar in unit of initial values. (b) and (e) show the $v_z-z$ phase space distributions at $0.067\ \si{ns}$ (labeled as dotted line in (a) and (d) respectively) with color-bar in arbitrary unit. (e) and (f) show the $v_z-z$ phase space distributions of shock front region (between the two dotted lines in (b) and (e) respectively) with color-bar in arbitrary unit.} 
\end{figure*}

Fig.\ \ref{fig4} shows the detailed simulation results for colliding velocity of $900\ \si{km/s}$ and initial center density of $1\ \si{g/cc}$ and $50\ \si{g/cc}$. It is evident that the shock front in Fig.\ \ref{fig4}(a) is weaker and blurred, which indicates that the thickness of shock front is compatible to the simulation scale. The thickness of shock is more discernible in the phase space. According to Fig.\ \ref{fig4}(b) and (e), the narrow transition region between clusters gathering around $v_z=\pm v_0$ to $v_z=0$, which is enlarged in Fig.\ \ref{fig4}(c) and (f), is clearly observed as the shock front of several $\si{\mu m}$s in length. Viewed along $v_z$ axis, particles in the slope of shock front violate the Maxwellian distribution, and the presumptions of local thermodynamic equilibrium no longer hold. Hence, it is necessary to step further surpassing the hydrodynamic theory, and investigate the kinetic effects in collidings with supreme high $M$.

Semi-quantitative kinetic analysis is conducted based on Mott Smith's and Tidman’s work \cite{MottSmith1951TheSO,Tidman1958}, in which the distribution function near the shock front is expressed by the superposition of two different equilibrium distributions

\begin{equation}
\begin{aligned}
  f(v,z)= &n_{\alpha }(z)(\frac{m_\text{D}}{2\pi k_B T_{\alpha }})^{3/2} \exp [-\frac{m_\text{D} (v-u_{\alpha })^2}{2k_BT_{\alpha }}]+\\
  &n_{\beta }(z)(\frac{m_\text{D}}{2\pi k_B T_{\beta }})^{3/2} \exp [-\frac{m_\text{D} (v-u_{\beta })^2}{2k_BT_{\beta }}],
  \label{eq5}
\end{aligned}
\end{equation}

\noindent where $u_{\alpha , \beta}$ is the average velocity. By substituting Eq.\ (\ref{eq5}) into the Fokker-Planck equation and performing integrals for velocity with stable condition $\partial f/\partial t=0$ , the number density in the shock front is deduced as

\begin{equation}
  \frac{n(z)}{n_0}=\frac{M^2+a-2+M^2 (a-1) e^{Bx/l}}{(M^2+a-2)(1+e^{Bx/l})}.
  \label{eq6}
\end{equation}

\noindent In Eq.\ (\ref{eq6}), $n_0=n(-\infty)$ is the number density of particles far away in front of the shock, which are considered unperturbed by the shock. The parameter $a$ is given by $a=2\gamma /(\gamma -1)$, $l$ is the mean free path, and $B$ is a coefficient depending on the collision model selected in the Fokker-Planck equation.

We further apply Eq.\ (\ref{eq6}) to our colliding shock problem. For monatomic D-ions, $\gamma = 5/3$ and $a=5$. In the supreme high Mach number limit $M^2 >> 1$, Eq.\ (\ref{eq6}) turns to

\begin{equation}
  \frac{n(z)}{n_0}=\frac{1+4 e^{Bz/l}}{1+e^{Bz/l}}=\frac{1+4 e^{z/\delta}}{1+e^{z/\delta}}.
  \label{eq7}
\end{equation}

\noindent where $l/B$ is on the scale of the shock thickness $\delta=[n(\infty)-n(-\infty)]/\vert dn/dz \vert _{max}$. Therefore, the structure of shock is dependent on the ratio of system spatial scale to shock thickness $z/\delta$. For inspection, in the hydrodynamic limit, as $z/\delta \to \infty$ for an infinitesimal shock thickness, according to Eq.\ (\ref{eq7}), $n(z)/n_0=4$, which is the maximum compression ratio in hydrodynamics.

In our one-dimensional collision cases, the spatial scale is of $\sim 100\ \si{\mu m}$, and the estimation of shock thickness refers to Keenan’s two-component analytical calculation \cite{Keenan2017pre}. The shock thickness is expressed as

\begin{equation}
  \delta=\delta_0 + \frac{m_i}{m_e}\lambda_\text{D}
  \label{eq8}
\end{equation}

\noindent where $\delta_0$ is proportional to $M^4$ (the theoretical derivation of this result is shown in Sec.\ \uppercase\expandafter{\romannumeral3} of \cite{SM}), and the last term is a kinetic modification merely as a function of dependent on the mean free path $\lambda_\text{D}=\frac{16\sqrt{6\pi} \epsilon_0^2 T^2}{n_\text{D}e^4ln\Lambda}$. According to Keenan’s model, for fully ionized plasmas, when $M>5$, the first term $\delta_0$ is starting to dominate over the second term. With colliding velocity of $900\ \si{km/s}$, density of $10\ \si{g/cc}$ and initial temperature of $50\ \si{eV}$, we have $M \approx 15$ and $\delta\sim 80\ \si{\mu m}$.

It is noted that the shock thickness $\delta$ is comparable to the length of the colliding region. According to Eq.\ (\ref{eq7}), the one-dimensional shock tube model has a steady-state downstream at $z \to \infty$, with a density four times of that in the upstream. However, in the colliding case, the profile of shock, which starts from the undisturbed upstream region, is cut off at the colliding center. As a result, the post-shock density of the center is apart from the $4$ times of compression supreme. Furthermore, as the colliding velocity increases, the Mach number of colliding shock accordingly increases, as well as the shock thickness shown in Eq.\ (\ref{eq8}). Since the compression ratio in Eq.\ (\ref{eq7}) increases monotonically with the coefficient $z/\delta$, the rise of $\delta$ decreases the final density in the colliding center, which sensitively accounts for the downtrend result of simulations shown in Fig.\ \ref{fig3}. In order to validate the above arguments, we have additionally performed another set of simulations for one-component ideal gases with interactions modelled by elastic sphere collisions. For fixed colliding velocity and gas density, the compression ratio is also decreasing when the shock thickness is increasing (Sec.\ \uppercase\expandafter{\romannumeral4} of Supplemental Material \cite{SM}).

In conclusion, we have investigated the effects of quantum degeneracy and kinetics in high Mach number collidings of two quantum degenerate plasmas. Via large-scale one-dimensional kinetic simulations and hydrodynamic calculations, we found both quantum degeneracy and kinetics play key roles in density compression. In order to achieve high-density compression, the colliding kinetic pressure should be significantly higher than the pressure by quantum degeneracy. However, when the Mach number is further increased, a decreasing trend of density compression is surprisingly observed, attributing to kinetic effects. This result is physically reasonable for plasmas. As the colliding velocity increases, the thickness of the shock is eventually comparable to the system scale. This means the shock is cut off in the middle by the colliding center and is apart from the 4 times compression supreme. Consequently, the shock structure significantly affects the physical properties in the post-shock region.

Our results provide a guide for the design of DCI experiments. It is suggested that a theoretically optimal colliding velocity can be found. At this velocity, the colliding kinetic pressure starts to surpass the degenerate pressure and the thickness of colliding shock is not significantly broadened by high-Mach number kinetics, resulting in theoretically highest density compression. Moreover, since the colliding of the degenerate plasmas is a common phenomenon in astrophysical systems, our results may be of relevance to the physics process such as the merger of neutron stars and white dwarfs.

\begin{acknowledgments}

W.-B. Zhang and Y.-H. Li contributed equally to this work. W.-B. Zhang conducted the one-dimensional simulations by the LAPINS code and contributed to the deduction of kinetic equations; Y.-H. Li was responsible for constructing the hydrodynamic colliding model and undertook a portion of the kinetic theoretical analysis. 

This work is supported by the Strategic Priority Research Program of Chinese Academy of Sciences (Grant Nos. XDA25010100 and XDA250050500), National Natural Science Foundation of China (Grants No. 12075204), and Shanghai Municipal Science and Technology Key Project (No. 22JC1401500). Dong Wu thanks the sponsorship from Yangyang Development Fund. 

\section{Supplementary materials}

\section{\expandafter{\romannumeral1}. Benchmark of The Simulation Setup}

To provide a benchmark of our colliding simulation configuration in the main text, we perform another set of large-scale kinetic simulations with the code LAPINS \cite{Wu2020PhysRevE,Wu2023}. To ensure an affordable time cost, the simulation region is set to $400\ \si{\mu m}$ while it is $1000\ \si{\mu m}$ in the main text. By keeping the total number of computational particles constant, we conduct simulations with different cell sizes.

\begin{figure}
	\includegraphics[width=8.5cm]{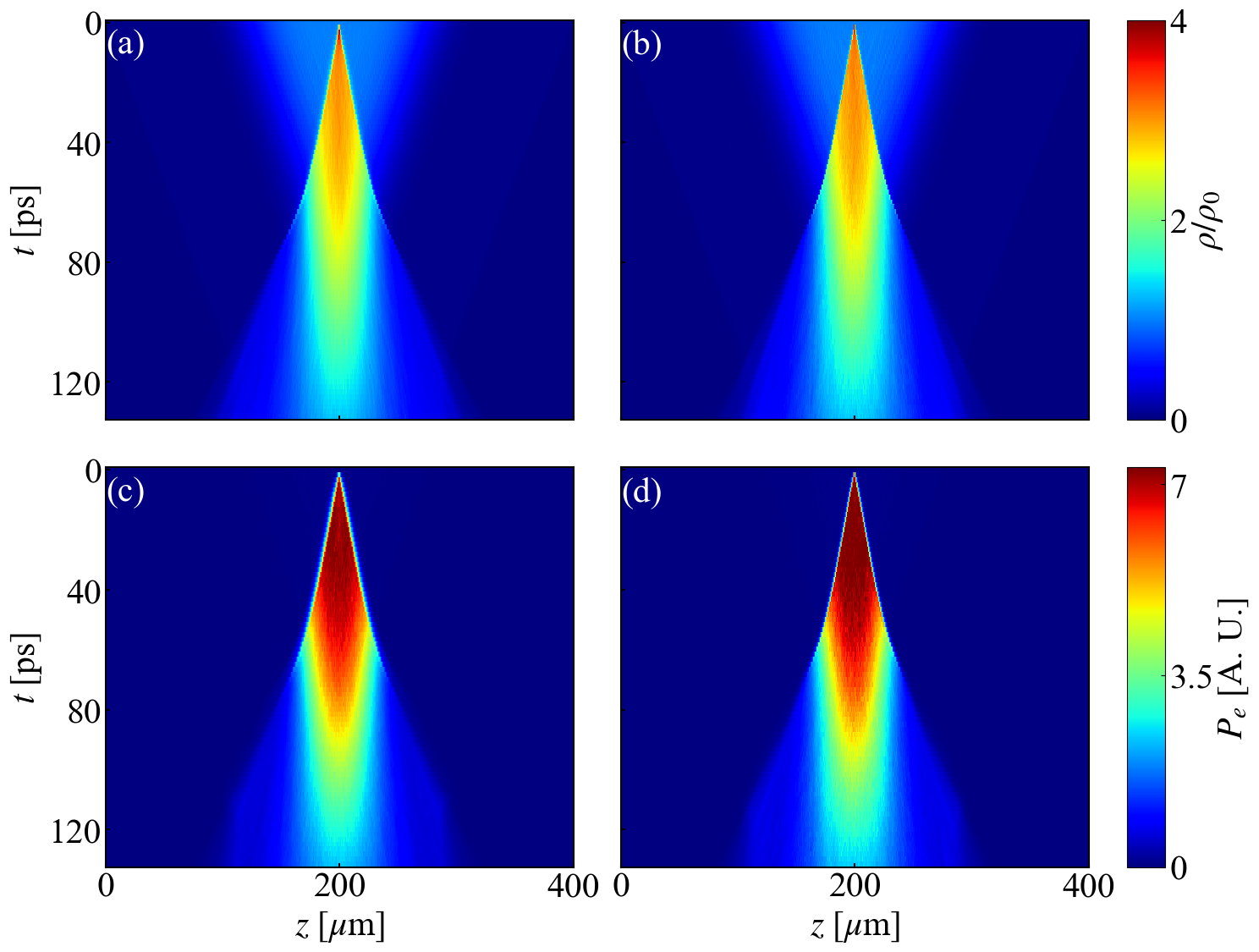} 
	\caption{\label{fig:S1}(color online). The spatial-temporal evolutions of plasma density (in first row) and electron pressure (in second row) when the colliding velocity is $v_0=500\ \si{km/s}$: (a) and (c) show the case when the cell size is $d_z=0.4\ \si{\mu m}$ and the number of particles per cell is $1000$. (b) and (d) show the case when the cell size is $d_z=0.2\ \si{\mu m}$ and the number of particles per cell is $500$. Values in the color-bar represent plasma density $\rho$ in unit of its initial value where $\rho_0=100$ g/cc and electron pressure $P_e$ in arbitrary unit.} 
\end{figure}

To illustrate the convergence of our simulations in detail, we first pay attention to the spatial-temporal evolutions of plasma density and electron pressure, which are shown in Fig.\ \ref{fig:S1}. The left column shows the case when the cell size is $d_z=0.4\ \si{\mu m}$ and the number of particles per cell is $1000$ and the right one shows the case when the cell size is $d_z=0.2\ \si{\mu m}$ and the number of particles per cell is $500$. Compared with the right column of Fig.\ \ref{fig:S1}, it is nearly identical in the left column in terms of the peak value and the profile of the physical quantities during the simulation time. It is therefore convergent in terms of the density and temperature evolutions during the simulation time.

We then focus on the temporal evolutions of the total energy of the electrons and the ions, which are shown in Fig.\ \ref{fig:S2} (the simulation parameters are the same as that in Fig.\ \ref{fig:S1}). Solid lines and dotted lines represent the cases when the cell size is $d_z=0.4\ \si{\mu m}$ and the cell size is $d_z=0.2\ \si{\mu m}$ respectively. Considering the simulations in the two cases, as displayed in Fig.\ \ref{fig:S2}, it is strictly convergent in the compression process which we are interested in and is nearly convergent in the diffusion process. Therefore, it is convergent in terms of the total energy evolutions during the simulation time.

To simulate the colliding plasma in a larger region with affordable time cost and reasonable nodes allocation, we slightly increase the cell size up to $d_z=0.5\ \si{\mu m}$ with the number of particles per cell $1000$ in the main text simulations. This may generate mere errors of the physical values, but it is justified in terms of the trend of the physical quantities.

\section{\expandafter{\romannumeral2}. Efficiency of Energy Conversion}

\begin{figure}
	\includegraphics[width=8.5cm]{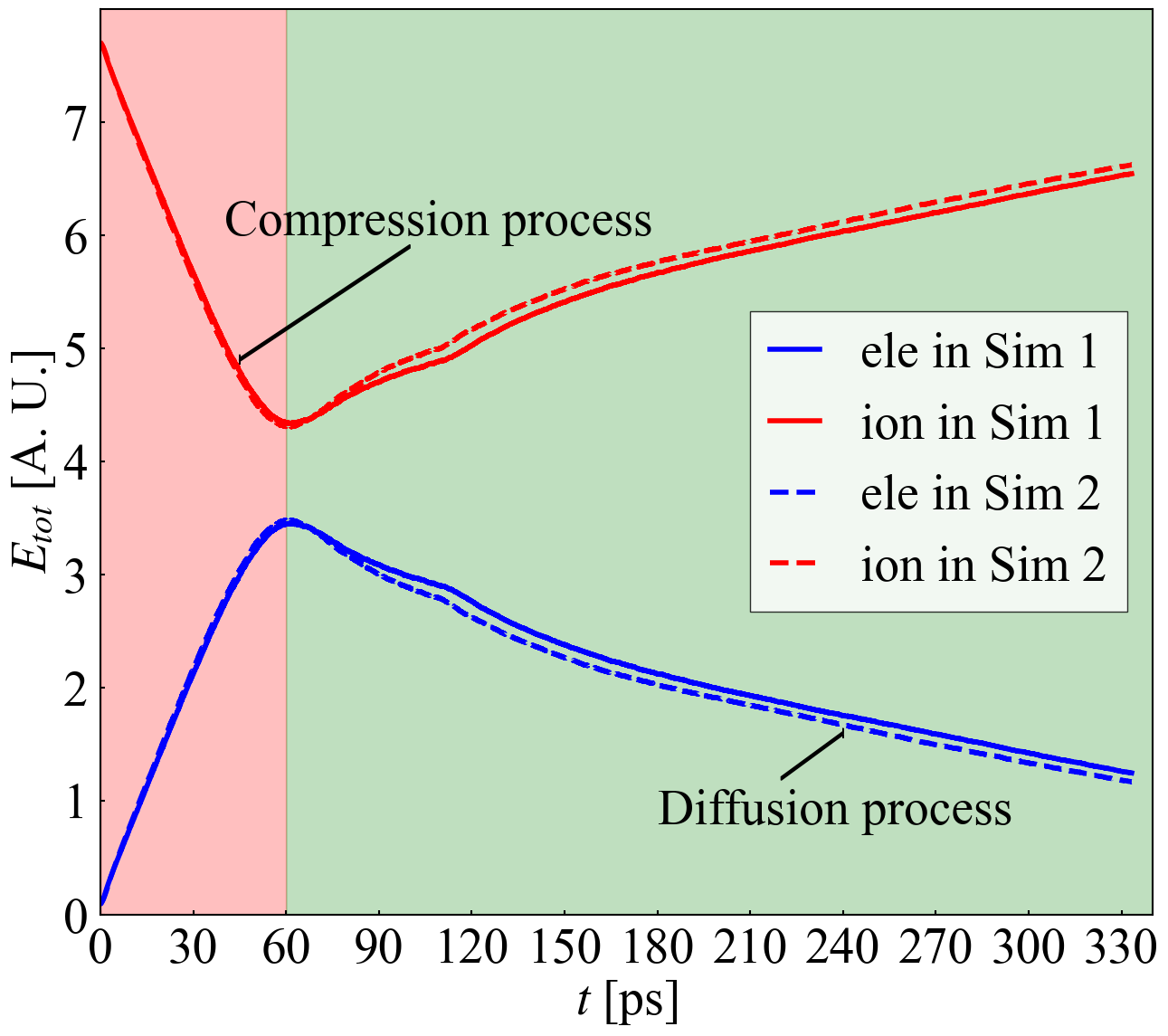} 
	\caption{\label{fig:S2}(color online). The temporal evolutions of the total energy of the electrons (blue lines) and the ions (red lines) when the colliding velocity is $v_0=500\ \si{km/s}$. Solid lines represent the results when the cell size is $d_z=0.4\ \si{\mu m}$ and the number of particles per cell is $1000$. Dotted lines represent the results when the cell size is $d_z=0.2\ \si{\mu m}$ and the number of particles per cell is $500$.} 
\end{figure}

In our hydrodynamic model, for the sake of simplicity, we assume that in the post-shock region, the drift kinetic energy of ions (the drift kinetic energy of electrons can be neglected in the simulation parameters) is entirely converted to the thermal energy of ions and electrons in our theoretical model. For the scenario of symmetric collision, we may directly conclude the zero drift velocity in the post-shock region, resulting in the above assumption. However, this assumption should be carefully confirmed.

The above assumption is verified by a series of large-scale kinetic simulations which were carried out. We pay attention to the temporal evolutions of the total energy of the electrons and the ions which are shown in Fig.\ \ref{fig:S2}. The convergence of the two cases is discussed in Sec.\ \uppercase\expandafter{\romannumeral1}, and here we focus on the total energy trend of the ions and electrons. In the compression process (red region in Fig.\ \ref{fig:S2}), as the drift velocity of ions and electrons decreases and the temperature of them increases, the kinetic energy is converted to the thermal energy. The mass of ions is far larger than that of electrons, so that the kinetic energy of ions is far larger than that of electrons when they have identical velocity. Additionally, the thermal energy of ions is the same as that of electrons when they have identical temperature. Accordingly, in the compression process, the total energy of ions decreases and that of electrons has a opposite trend. For the same reason, the total energy of ions increases and that of electrons decreases in the diffusion region. At the junction of the red and green regions, since the total energy of ions and electrons are close, the majority of kinetic energy is converted to the thermal energy, which indicates the rationality of our theoretical assumption.

\section{\expandafter{\romannumeral3}. Kinetic Theory for Strong Shock Thickness in Plasmas}

In the colliding process of high Mach number plasmas, the key reason for the decreasing trend of density compression is that the thickness of shock scales as $K^4$ where $K$ is the Mach number. This is the result derived by Mott-Smith and Tidman \cite{MottSmith1951TheSO,Tidman1958} theoretically. In the following parts of this section, we briefly summarize the derivation.

For simplify, we consider plasmas with protons and electrons moving in the $x$ direction, which have masses $M$ and $m$ per particle respectively. Consistent with Mott-Smith treatment, the distribution of ion $F$ is bi-Maxwell, namely

\begin{equation}
	\begin{aligned}
		F= & N_\alpha(x)\left(\frac{M}{2 \pi k T_\alpha}\right)^{\frac{3}{2}} \exp \left[-\frac{M}{2 k T_\alpha}\left(\mathbf{c}-\mathbf{i} U_\alpha\right)^2\right] \\
		& +N_\beta(x)\left(\frac{M}{2 \pi k T_\beta}\right)^{\frac{3}{2}} \exp \left[-\frac{M}{2 k T_\beta}\left(\mathbf{c}-\mathbf{i} U_\beta\right)^2\right] \\
		= & F_\alpha+F_\beta,
		\label{eq1}
	\end{aligned}
\end{equation}

\noindent where the suffix $\alpha$ and $\beta$ represent the conditions ahead of and behind the shock respectively, $\mathbf{c}$ is velocity, $N_{\alpha}$ and $N_{\beta}$ are densities, $T_{\alpha}$ and $T_{\beta}$ are temperatures and $U_{\alpha}$ and $U_{\beta}$ are stream velocities along the $x$ direction $\mathbf{i}$. Since the relaxation time for electrons to reach equilibrium is small, we take electron distribution $f$ as self-equilibrium, namely

\begin{equation}
	f(x)=n(x)\left(\frac{m}{2 \pi k T(x)}\right)^{\frac{3}{2}} \exp \left[-\frac{m}{2 k T}\left(\mathbf{c}-\mathbf{i} U_e\right)^2\right],
	\label{eq2}
\end{equation}

\noindent where $n$, $T$ and $U_e$ are all functions of $x$.

When simplified for two-body Coulomb interactions, the evolutions of the two distribution functions $F$ and $f$ are described by Fokker-Planck equations, namely

\begin{equation}
	\begin{aligned}
		& \frac{\partial F}{\partial t}+\mathbf{c} \cdot \frac{\partial F}{\partial \mathbf{r}}+\frac{e \mathbf{E}}{M} \cdot \frac{\partial F}{\partial \mathbf{c}}=\left(\frac{\partial F}{\partial t}\right)_c, \\
		& \frac{\partial f}{\partial t}+\mathbf{c} \cdot \frac{\partial f}{\partial \mathbf{r}}-\frac{e \mathbf{E}}{m} \cdot \frac{\partial f}{\partial \mathbf{c}}=\left(\frac{\partial f}{\partial t}\right)_c,
		\label{eq3}
	\end{aligned}
\end{equation}

\noindent where the collision terms are given by

\begin{equation}
	\begin{aligned}
		& \frac{1}{\Gamma}\left(\frac{\partial F}{\partial t}\right)_c=-\frac{\partial}{\partial c_i}\left(F \frac{\partial H}{\partial c_i}\right)+\frac{1}{2} \frac{\partial}{\partial c_i} \frac{\partial}{\partial c_j}\left(F \frac{\partial}{\partial c_i} \frac{\partial}{\partial c_j} G\right), \\
		& \frac{1}{\gamma}\left(\frac{\partial f}{\partial t}\right)_c=-\frac{\partial}{\partial c_i}\left(f \frac{\partial h}{\partial c_i}\right)+\frac{1}{2} \frac{\partial}{\partial c_i} \frac{\partial}{\partial c_j}\left(f \frac{\partial}{\partial c_i} \frac{\partial}{\partial c_j} g\right),
		\label{eq4}
	\end{aligned}
\end{equation}

\noindent and

\begin{equation}
	\begin{aligned}
		& h=2 \int d \mathbf{c}_1 \frac{f\left(\mathbf{c}_1\right)}{\left|\mathbf{c}-\mathbf{c}_1\right|}+\left(\frac{m+M}{M}\right) \int d \mathbf{c}_1 \frac{F\left(\mathbf{c}_1\right)}{\left|\mathbf{c}-\mathbf{c}_1\right|}, \\
		& g=G=\int d \mathbf{c}_1(f+F)\left|\mathbf{c}-\mathbf{c}_1\right|, \\
		& H=\left(\frac{M+m}{m}\right) \int d \mathbf{c}_1 \frac{f\left(\mathbf{c}_1\right)}{\left|\mathbf{c}-\mathbf{c}_1\right|}+2 \int d \mathbf{c}_1 \frac{F\left(\mathbf{c}_1\right)}{\left|\mathbf{c}-\mathbf{c}_1\right|} .
		\label{eq5}
	\end{aligned}
\end{equation}

\noindent The slowly varying quantity $\Gamma$ is

\begin{equation}
	\Gamma=\frac{4 \pi e^4}{M^2} \ln \left[\frac{3}{4(\pi n)^{\frac{1}{2}}}\left(\frac{k T}{e^2}\right)^{\frac{3}{2}}\right],
	\label{eq6}
\end{equation}

\noindent and $\gamma$ is obtained by replacing $M$ with $m$ in Eq.\ (\ref{eq6}). The form of collision terms used is the same as that used by Rosenbluth, MacDonald and Judd \cite{Rosenbluth1957}.

Multiplying Eq.\ (\ref{eq3}) by $v^2$ ($\mathbf{c}=(u,v,w)$) and integrating the collision terms Eq.\  (\ref{eq4})-(\ref{eq5}) over $\mathbf{c}$ by parts using $F(c=\pm \infty)= f(c=\pm \infty)=0$, we find

\begin{equation}
	\begin{aligned}
		& U_\alpha\left(\frac{k T_\alpha}{M}\right) \frac{\partial N_\alpha}{\partial x}+U_\beta\left(\frac{k T_\beta}{M}\right) \frac{\partial N_\beta}{\partial x}+\frac{U_\beta N_\beta k}{M} \frac{\partial T_\beta}{\partial x} \\
		& \quad=2\left(\frac{M+m}{m}\right) \Gamma \int d \mathbf{c} v F(\mathbf{c}) \int d \mathbf{c}_1 f\left(\mathbf{c}_1\right) \frac{\partial}{\partial v}\left|\mathbf{c}-\mathbf{c}_1\right|^{-1} \\
		& \quad+\Gamma \int d \mathbf{c} F(\mathbf{c}) \int d \mathbf{c}_1 F\left(\mathbf{c}_1\right) \frac{\partial^2}{\partial v^2}\left|\mathbf{c}-\mathbf{c}_1\right| \\
		& \quad+4 \Gamma \int d \mathbf{c} v F(\mathbf{c}) \int d \mathbf{c}_1 F\left(\mathbf{c}_1\right) \frac{\partial}{\partial v}\left|\mathbf{c}-\mathbf{c}_1\right|^{-1} \\
		& \quad+\Gamma \int d \mathbf{c} f(\mathbf{c}) \int d \mathbf{c}_1 F\left(\mathbf{c}_1\right) \frac{\partial^2}{\partial v^2}\left|\mathbf{c}-\mathbf{c}_1\right|,
		\label{eq7}
	\end{aligned}
\end{equation}

\noindent for proton $v^2$ equation. In the shock region, since $T_{\beta}$ has a slow variation compared with $N_{\alpha}(x)$ and $N_{\beta}(x)$ which we are interested in, it can be considered as constant. Thus Eq.\ (\ref{eq7}) becomes

\begin{equation}
	\begin{aligned}
		U_\alpha\left(\frac{k T_\alpha}{M}\right) \frac{\partial N_\alpha}{\partial x}+&U_\beta\left(\frac{k T_\beta}{M}\right)  \frac{\partial N_\beta}{\partial x} \\
		& =\frac{2 \Gamma N_\alpha N_\beta}{\pi^3}\left(\frac{M}{2 k T_\alpha}\right)^{\frac{1}{2}} \Psi ,
		\label{eq8}
	\end{aligned}
\end{equation}

\noindent where

\begin{equation}
	\begin{aligned}
		\Psi= & \left(\frac{2 k T_\alpha}{M}\right)^{\frac{1}{2}} \int d \mathbf{c} d \mathbf{c}_1 \exp \left[-\left(c^2+c_1^2\right)\right] \\
		& \times\left\{\left|\left(\frac{2 k T_\alpha}{M}\right)^{\frac{1}{2}} \mathbf{c}-\left(\frac{2 k T_\beta}{M}\right)^{\frac{1}{2}} \mathbf{c}_1+\mathbf{i}\left(U_\alpha-U_\beta\right)\right|^{-1}\right. \\
		- & \left.\frac{3\left[\left(2 k T_\alpha / M\right)^{\frac{1}{2}} v-\left(2 k T_\beta / M\right)^{\frac{1}{2}} v_1\right]^2}{\left|\left(2 k T_\alpha / M\right)^{\frac{1}{2}} \mathbf{c}-\left(2 k T_\beta / M\right)^{\frac{1}{2}} \mathbf{c}_1+\mathbf{i}\left(U_\alpha-U_\beta\right)\right|^3}\right\} .
		\label{eq9}
	\end{aligned}
\end{equation}

\noindent Notice that $\Psi$ can be evaluated by using Fourier integrals.

We next make use of the conservation of mass equation, namely

\begin{equation}
	\begin{gathered}
		N_\alpha U_\alpha+N_\beta U_\beta=\bar{N}_\alpha U_\alpha , \\
		\frac{\partial N_\beta}{\partial x}=-\frac{U_\alpha}{U_\beta} \frac{\partial N_\alpha}{\partial x},
		\label{eq10}
	\end{gathered}
\end{equation}

\noindent where $\bar{N}_{\alpha}= N_{\alpha}(-\infty)$. Therefore, Eq.\ (\ref{eq8}) becomes

\begin{equation}
	\begin{aligned}
		\frac{\partial N_\alpha}{\partial x}&\left[\frac{1}{N_\alpha}+\frac{1}{\bar{N}_\alpha-N_\alpha}\right] \\
		&=\frac{2 \Gamma \bar{N}_\alpha}{\pi^3 U_\beta}\left(\frac{M}{2 k T_\alpha}\right)^{\frac{1}{2}} \frac{M \Psi}{k\left(T_\alpha-T_\beta\right)} .
	\end{aligned}
	\label{eq11}
\end{equation}

\noindent By choosing the origin $x=0$ at the point where $ N_{\alpha}(x=0)=\frac{1}{2}\bar{N}_{\alpha}$, we will find the solutions of Eq.\ (\ref{eq11}) which is formally similar to that in Mott-Smith work \cite{MottSmith1951TheSO}, namely

\begin{equation}
	\begin{aligned}
		& N_\alpha=\frac{\bar{N}_\alpha e^{-x / l}}{\left(1+e^{-x / l}\right)}, \\
		& N_\beta=\frac{\bar{N}_\beta}{\left(1+e^{-x / l}\right)},
		\label{eq12}
	\end{aligned}
\end{equation}

\noindent where the shock thickness $l$ is

\begin{equation}
	l=\frac{\pi^3 U_\beta k\left(T_\beta-T_\alpha\right)}{2 \Gamma \bar{N}_\alpha M \Psi}\left(\frac{2 k T_\alpha}{M}\right)^{\frac{1}{2}}.
	\label{eq13}
\end{equation}

Introducing Mach number $K=(U_{\alpha}/V)$ for the stream ahead of the shock where $V$ is the velocity of sound in the plasma, we can express the shock thickness as

\begin{equation}
	\begin{aligned}
		l\left(\frac{\bar{N}_\alpha \ln \Lambda}{V^4}\right)=&\frac{3 \pi^2}{128}\left(\frac{3}{5}\right)^{\frac{1}{2}} \frac{M^2 K\left(3+K^2\right)}{e^4 \Psi} \\
		&\times\left(\frac{1}{4}-\frac{3}{20 K^4}-\frac{1}{10 K^2}\right).
	\end{aligned}
	\label{eq14}
\end{equation}

\noindent For $K$ is large,

\begin{equation}
	\Psi \rightarrow 0.309 \pi^4 / a ,
	\label{eq15}
\end{equation}

\noindent so that

\begin{equation}
	l \rightarrow \frac{29.1 K^4 V^4}{512 \pi \bar{N}_\alpha \Gamma},
	\label{eq16}
\end{equation}

\noindent which indeed reveals that the shock thickness scales as $K^4$ as $K$ is large.

\section{\expandafter{\romannumeral4}. The Colliding of One-component Ideal Gasses with Elastic Sphere Collisions}

To verify the robustness of our kinetic interpretation in the main text, another set of large-scale simulations are carried out. In these simulations, the simulated particles are modelled as elastic spheres instead of charged particle, so that the thickness of the shock is proportional to the reciprocal differential cross section \cite{MottSmith1951TheSO}, namely

\begin{equation}
	l \sim 1/\sigma^2
	\label{eq4-1}
\end{equation}

where $l$ is the shock thickness and $\sigma$ is the diameter of the elastic sphere. In the LAPINS code, the characteristic velocity $v_r$, which is in unit of $c$ the speed of light, is introduced to adjust the diameter $\sigma$ with the relation

\begin{equation}
	\frac{1}{2}mv_r^2 = \frac{1}{4\pi \epsilon_0}\frac{e^2}{\sigma}
	\label{eq4-2}
\end{equation}

where $m$ is the mass of a single deuterium particle. Accordingly, the shock thickness $l$ scales as $v_{r}^{4}$, so that it is a simpler scenario to verify our interpretation in the main text.

\begin{figure}
	\includegraphics[width=8.5cm]{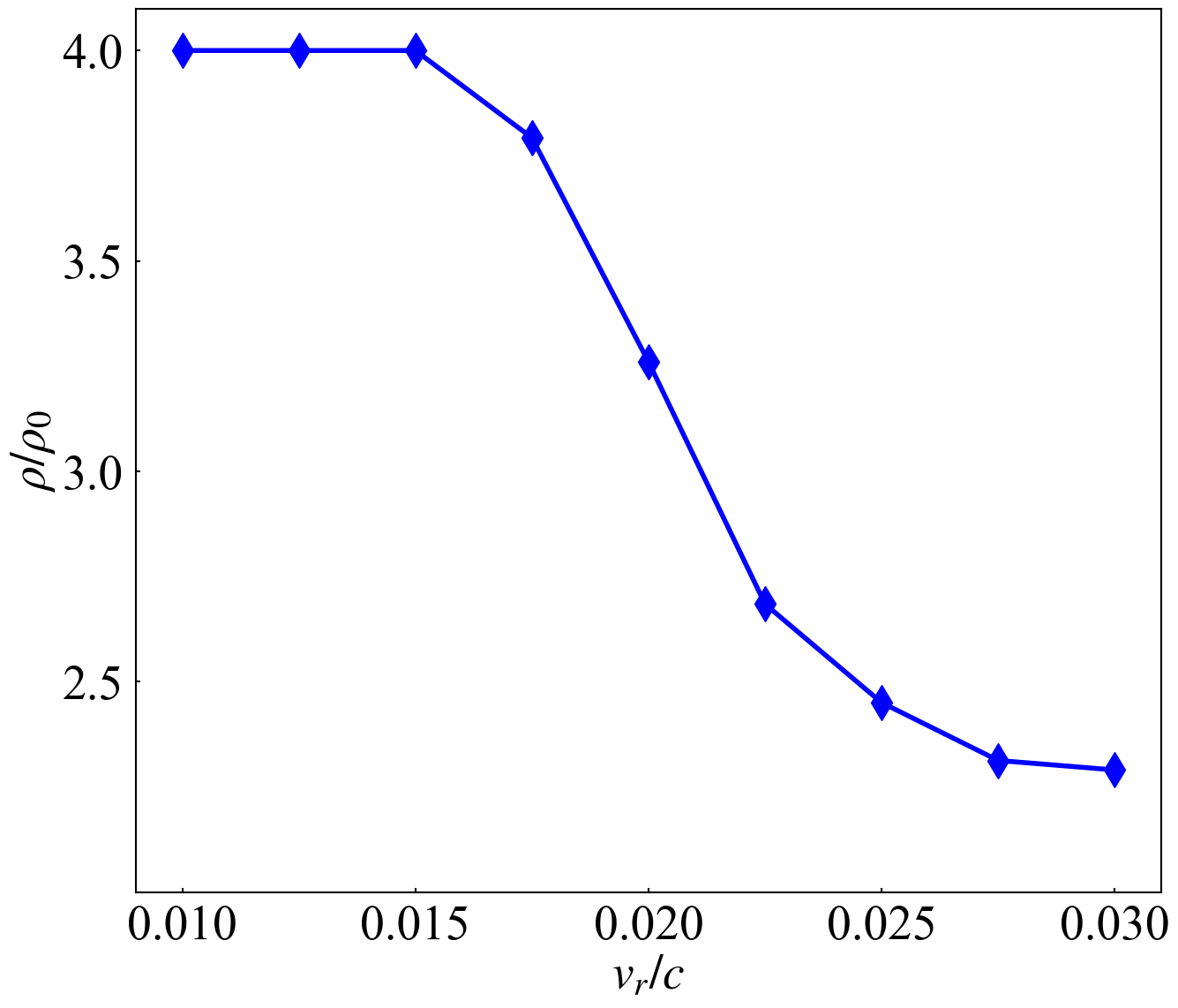} 
	\caption{\label{fig:S3}(color online). Density of the post-shock region, in unit of its initial value, where $\rho_0=100\ \si{g/cc}$. $v_r$ is the characteristic velocity in unit of $c$ the speed of light, which is relevant to the differential cross section. The colliding velocity is $v_0=500\ \si{km/s}$.} 
\end{figure}

The density compression results are shown in Fig.\ \ref{fig:S3}. It should be remarked that the colliding velocity is initialized as $v_0=500\ \si{km/s}$, which is large enough to assure the supreme density compression ratio $4$. In addition, the initial density is $\rho_0=100\ \si{g/cc}$, which is the same as that in the main text simulations. As shown in Fig.\ \ref{fig:S3}, as the characteristic velocity $v_r$ increases, the post-shock density decreases from $4$ to nearly $2$. Since the shock thickness $l$ scales as $v_{r}^{4}$, this result is consistent with the interpretation in the main text.

\end{acknowledgments}

\bibliography{apssamp}

\end{document}